\documentclass[]{spie}  %>>> use for US letter paper
\usepackage{amsmath,amsfonts,amssymb}
\usepackage{graphicx}
\usepackage[colorlinks=true, allcolors=blue]{hyperref}
\usepackage{multirow}
\usepackage{tikz-cd} % commutative diagrams
\usepackage{tikz}
\usepackage{comment}
\usetikzlibrary{graphs,graphs.standard,quotes,arrows,calc,matrix}
\usepackage{stmaryrd}
\usepackage{color}
\tikzset{
  LabelStyle/.style = { rectangle, rounded corners, draw,
                        minimum width = 2em, fill = yellow!50,
                        text = red, font = \bfseries },
  VertexStyle/.append style = { inner sep=5pt,
                                font = \normalsize\bfseries},
  EdgeStyle/.append style = {->, bend left} }
\usetikzlibrary {positioning}
\title{On the application of Sylvester's law of inertia to QUBO formulations for systems of linear equations}

\author[a]{Sun Woo Park}
\author[b,*]{Kyungtaek Jun}
\affil[a]{Department of Mathematics, University of Wisconsin-Madison, 480 Linconln Dr., Madison, WI 53706, USA / National Institute for Mathematical Sciences, 463-1 Jeonmin-dong, Yuseong-gu, Daejon, 34047, Republic of Korea.}
\affil[b]{Research Center, Innovation on Quantum and Computed Tomography, Seoul, Republic of Korea}

\authorinfo{
Corresponding Author: Kyungtaek Jun, \\
Sun Woo Park: E-mail: spark483@wisc.edu / spark483@nims.re.kr, \\ 
Kyungtaek Jun: E-mail: ktfriends@gmail.com}

\begin{document} 
\maketitle

\begin{abstract}
Previous research on quantum annealing methods focused on effectively modeling systems of linear equations by utilizing quadratic unconstrained binary optimization (QUBO) formulations. These studies take part in enhancing quantum computing algorithms, which extract properties of quantum computers suitable for improving classical computational models. In this paper, we further develop the QUBO formulations of systems of linear equations by applying Sylvester's law of inertia, which explores matrix congruence of any real symmetric matrix to a diagonal matrix. We expect that the proposed algorithm can effectively implement higher dimensional systems of linear equations on a quantum computer. In particular, the proposed algorithm hints the linear correspondence between the number of unknown variables of a linear system and the number of qubits supported on a quantum annealing device. Further experimental verification of the proposed QUBO models as well as their comparisons to classical algorithms are also made.
\end{abstract}


\section{INTRODUCTION}
\label{sec:intro}  % \label{} allows reference to this section

Recent research on quantum computers focus on devising an effective and efficient algorithm which surpasses or complements previously studied algorithms. Among the newly proposed quantum algorithms, quantum annealing method focuses on devising an optimization problem which minimizes an energy level function induced from pre-existent classical models \cite{RI14, MV16}. The problem of solving a system of linear equations, which is the focus of this paper, can be also reformulated as an energy minimization problem. In particular, quadratic unconstrained binary optimization (QUBO) formulations, proposed by Borle and Lomonaco, can be utilized to constructing a quadratic energy function for solving a system of linear equations \cite{BL19}. Quantum annealing processors effectively parallelize the procedure of solving the minimization problem by utilizing the base 2 representation of real numbers \cite{MV16, BL19}. This has potential to greatly reduce computational complexity in solving higher dimensional systems of linear equations compared to previously studied methods. However, physical restraints of contemporary quantum computers may hinder such potential of QUBO formulations. These include partial connectivity of qubits and the limited number of implementable qubits on a quantum computer \cite{BL19, JC21}. It is therefore a crucial question to ask what quantum algorithmic simplifications are required to surpass classical algorithms in solving a system of linear equations with large number of variables.

Recent research on such QUBO formulations focuses on utilizing matrix congruence to obtain a new simplified model for solving linear systems of equations \cite{PLKWJ21}. This congruence relation is known as ``Sylvester's law of inertia", which guarantees the transformations of real symmetric matrices to diagonal matrices, regardless of whether the given symmetric matrices are invertible or not \cite{Ay62}. We further examine the proposed methodology to show that Sylveter's law can significantly extend the implementable number of variables of a system of linear equations on a quantum computer. This feature is obtained from the fact that the proposed QUBO formulation substantially decreases the occurrences of required entanglements among utilized qubits, which reduces the required qubit connectivity of quantum computers. 

We expect that the proposed feasible optimization technique opens up numerous possibilities to effectively and efficiently implement a wide range of systems of linear equations on a quantum annealing device. One particular merit of the proposed method, which we will primarily focus on this paper, is that Sylvester's law of inertia hints the linear correspondence between the number of unknown variables of a linear system and the number of qubits supported on a quantum annealing device. For instance, we conjecture that the upcoming D-wave 7000Q quantum annealing device can process systems of linear equations with $\frac{7}{2}$ times more unknown variables than those implementable on the D-wave 2000Q quantum computer. Careful constructions and examinations of some examples are provided for verification of the potential effectiveness, efficiency, and extensiveness of the proposed algorithm. 

%\textcolor{red}{I added a comment. Please check the source.}
%초록과 서론에 "Quantum annealers에서 한 문제에 사용할 수 있는 총 큐빗 수와 선형방정식의 크기가 선형적으로 비례하는 QUBO 모델링이 가능하다. 이것은 D-Wave 7000Q에서 20비트 수준의 정밀한 해를 구할 수 있다는 것을 의미한다"는 내용을 추가 하는게 어떨까요?, 제 생각에는 우리의 장점을 서론과 초록에 잘 적어 넣는게 중요한거 같아요. (check - 박선우)
%두번째 단락 시작시에 사람 이름은 2명까지 넣을 수 있어요. Previous research나 Park et al로 가시면 될거 같아요 (check - 박선우)
%Scientific Reports나 Nature communications 버젼으로 바꿔 주실 수 있으세요?
%지금은 서로 의견을 나누는 단계니까... 선우씨 의견도 마음껏 말씀해 주세요.
%제가 요새 글을 좀만 보면 어질어질해서 오래 못봅니다 ㅠㅠ, 최대한 열심히 해볼께요.

\section{Method}
\label{sec: method}

\subsection{Sylvester's Law of Inertia}
\label{subsec: background}

We recall how Sylvester's Law of Inertia simplifies the QUBO formulation of systems of linear equations \cite{PLKWJ21}. Let $A = (a_{i,j})_{i,j=1}^n \in \mathbb{R}^{n \times n}$ be an invertible real matrix, and $b = (b_i)_{i=1}^n \in \mathbb{R}^n$ a real column vector. We denote the vector of $n$ unknown variables as $x = (x_i)_{i=1}^n \in \mathbb{R}^n$. Suppose we have the system of linear equations given by
\begin{equation} \label{equation:linear_system}
    Ax = b.
\end{equation}
The solution which minimizes the $l^2$-norm of the equation
\begin{equation} \label{equation:minimize_square}
    \| Ax - b \|^2 = x^T A^T A x - 2b^T A x + b b^T
\end{equation}
solves (\ref{equation:linear_system}).

Using the $l^2$-norm, we define the energy level function:
\begin{equation} \label{equation:energy_function_defn}
    f(x) = x^T A^T A x - 2b^T A x.
\end{equation}

Then the $l^2$-norm minimizing solution of \eqref{equation:minimize_square} is the solution $x^* \in \mathbb{R}^n$ which satisfies
\begin{equation}
    f(x^*) = -b b^T.
\end{equation}

%\textcolor{red}{I added a comment. Please check the source.}
% 솔루션이 존재할 때 x^*이 존재해서 f(x^*) = -bb^T 이런식으로 전개하면 어떨까요? 변수 x말구용! (check-SWP)
% 근데 우리 이번 논문은 invertible할때만 하니까... 역행렬이 존재 할 때라고 하는게 더 좋을거 같아요. (check-SWP)

Recall that $A^T A$ is a positive semidefinite symmetric matrix over $\mathbb{R}$. Sylvester's law of inertia, which is stated below, implies that $A^T A$ can be transformed into a diagonal matrix $D \in \mathbb{R}^{n \times n}$ using matrix congruence relations.

\begin{theorem}[Sylvester's Law of Inertia \cite{S52}]
Let $S \in \mathbb{R}^{n \times n}$ be any real symmetric matrix. Then there exists a diagonal matrix $D = (d_{i,j})_{i,j=1}^n \in \mathbb{R}^{n \times n}$ and a non-singular real matrix $R= (r_{i,j})_{i,j=1}^n \in \mathbb{R}^{n \times n}$ such that 
\begin{equation} \label{equation:sylvester}
    D = R^T S R.
\end{equation}
Furthermore, the number of non-zero diagonal entries of $D$ is equal to the rank of the matrix $S$.
\end{theorem}

Using (\ref{equation:sylvester}), the equation (\ref{equation:minimize_square}) can be written as:
\begin{equation} \label{equation:sylvester_1}
    \| Ax - b \|^2 = (R^{-1} x)^T R^T A^T A R  (R^{-1} x) - 2b^T A R (R^{-1} x) + b b^T
\end{equation}
Let $y \in \mathbb{R}^n$ be a new vector of $n$ unknown variables defined as
\begin{equation}
    y = R^{-1} x.
\end{equation}
Note that $R^{-1}$ is a well-defined matrix because $R$ is non-singular.
Using the new variable $y$, we can simplify (\ref{equation:sylvester_1}) as:
\begin{equation} \label{equation:sylvester_2}
    \|Ax - b\|^2 = y^T D y - 2 (b^T A R) y + b b^T.
\end{equation}
The solution which optimizes the $l^2$-norm of (\ref{equation:sylvester_2}) gives the solution to the linear system (\ref{equation:linear_system}). 
\begin{align} \label{equation:sylvester_final}
    \|Ax - b\|^2 &= \sum_{i=1}^n d_{i,i} y_i^2 - 2 \sum_{i=1}^n \sum_{j=1}^n \sum_{k=1}^n  b_k a_{k,j} r_{j,i} y_i + \sum_{i=1}^n b_i^2
\end{align}
The energy function from \eqref{equation:energy_function_defn} can be reformulated as
\begin{equation} \label{equation:energy_function_simplify}
    f(y) = y^T D y - 2 b^T A R y = \sum_{i=1}^n d_{i,i} y_i^2 - 2 \sum_{i=1}^n \sum_{j=1}^n \sum_{k=1}^n  b_k a_{k,j} r_{j,i} y_i.
\end{equation}

\subsection{QUBO models}
Quantum annealing method can be used to approximate the $l^2$-norm minimizing solution of (\ref{equation:sylvester_2}) by utilizing a combination of qubits $q_{i,j} \in \{0,1\}$ \cite{MV16, BL19, JC21, J21, PLKWJ21}. In this subsection, we demonstrate that the proposed QUBO model greatly reduces the complexity of the energy function associated to system of linear equations, compared to that obtained from previously studied quantum annealing approaches.

The number of qubits required for implementing the QUBO model depends on the choice of the representations of variables. In this manuscript, we consider two types of representations, as proposed by O'Malley and Vesselinov \cite{MV16}, and Borle and Lomonaco \cite{BL19}.

We start with the base-2 representation of the column vector $y$ \cite{MV16}:
\begin{equation} \label{equation:QUBO_model_1}
    y_i \approx \sum_{l=-m}^m 2^l q_{i,l}^+ - \sum_{l=-m}^m 2^l q_{i,l}^-
\end{equation}
The above formulation allows $y_i$ to take both positive and negative values. In fact, the set of qubits satisfy the condition that for any digits $-m \leq l_1, l_2 \leq m$, \cite{J21, PLKWJ21}
\begin{equation} \label{equation:qubit_product}
    q_{i,l_1}^+ q_{i,l_2}^- = 0.
\end{equation}

Under this choice of representation, we can approximate (\ref{equation:energy_function_simplify}) as follows. The first term of (\ref{equation:energy_function_simplify}) reduces to:
\begin{align}
\begin{split}
    \sum_{i=1}^n d_{i,i} y_i^2 &\approx \sum_{i=1}^n d_{i,i} \left( \sum_{l=-m}^m 2^l q_{i,l}^+ - \sum_{l=-m}^m 2^l q_{i,l}^- \right)^2 \\
    &= \sum_{i=1}^n d_{i,i} \left( \left( \sum_{l=-m}^m 2^l q_{i,l}^+ \right)^2 + \left( \sum_{l=-m}^m 2^l q_{i,l}^- \right)^2\right) \\
    &= \sum_{i=1}^n d_{i,i} \left( \sum_{l=-m}^m 2^{2l} \left( \left(q_{i,l}^+ \right)^2 + \left( q_{i,l}^- \right)^2 \right)+ \sum_{l_1 < l_2} 2^{l_1 + l_2 + 1} \left( q_{i,l_1}^+ q_{i,l_2}^+ + q_{i,l_1}^- q_{i,l_2}^- \right) \right) \\
    &= \sum_{i=1}^n d_{i,i} \left( \sum_{l=-m}^m 2^{2l} \left( q_{i,l}^+ + q_{i,l}^- \right) + \sum_{l_1 < l_2} 2^{l_1 + l_2 + 1} \left( q_{i,l_1}^+ q_{i,l_2}^+ + q_{i,l_1}^- q_{i,l_2}^- \right) \right) \label{equation:QUBO_model_1_1}
\end{split}
\end{align}

The second term of (\ref{equation:energy_function_simplify}) reduces to:
\begin{align} \label{equation:QUBO_model_1_2}
    2 \sum_{i=1}^n \sum_{j=1}^n \sum_{k=1}^n  b_k a_{k,j} r_{j,i} y_i &\approx \sum_{i=1}^n \sum_{j=1}^n \sum_{k=1}^n \sum_{l=-m}^m 2^{l+1} b_k a_{k,j} r_{j,i} \left( q_{i,l}^+ - q_{i,l}^- \right).
\end{align}
The associated QUBO model is the summation of (\ref{equation:QUBO_model_1_1}) and (\ref{equation:QUBO_model_1_2}), i.e.
\begin{align}
    \begin{split}
        f(y) &= \sum_{i=1}^n d_{i,i} \left( \sum_{l=-m}^m 2^{2l} \left( q_{i,l}^+ + q_{i,l}^- \right) + \sum_{l_1 < l_2} 2^{l_1 + l_2 + 1} \left( q_{i,l_1}^+ q_{i,l_2}^+ + q_{i,l_1}^- q_{i,l_2}^- \right) \right) \\
        &+ \sum_{i=1}^n \sum_{j=1}^n \sum_{k=1}^n \sum_{l=-m}^m 2^{l+1} b_k a_{k,j} r_{j,i} \left( q_{i,l}^+ - q_{i,l}^- \right)
    \end{split}
\end{align}

Borle and Lomonaco's approximation \cite{BL19} of $y$, on the other hand, takes the following form:
\begin{equation} \label{equation:QUBO_model_2}
    y_i \simeq -2^{m+1} q_i^- + \sum_{l=-m}^m 2^l q_{i,l}^+
\end{equation}
This approach leads to a new QUBO model which further reduces the number of utilized qubits. The first term of (\ref{equation:sylvester_final}) is approximated by:
\begin{align} 
\begin{split}
    \sum_{i=1}^n d_{i,i} y_i^2 &\simeq \sum_{i=1}^n d_{i,i} \left( -2^{m+1} q_i^- + \sum_{l=-m}^m 2^l q_{i,l}^+ \right)^2 \\
    &= \sum_{i=1}^n d_{i,i} \left( 2^{2m+2} (q_i^-)^2 + \sum_{l=-m}^m \left( 2^{2l} (q_{i,l}^+)^2 - 2^{m+l+2} q_i^- q_{i,l}^+  \right) +  \sum_{l_1 < l_2} 2^{l_1 + l_2 + 1} q_{i,l_1}^+ q_{i,l_2}^+ \right) \\
    &= \sum_{i=1}^n d_{i,i} \left( 2^{2m+2} q_i^- + \sum_{l=-m}^m 2^{2l} q_{i,l}^+ - \sum_{l=-m}^m 2^{m+l+2} q_i^- q_{i,l}^+ + \sum_{l_1 < l_2} 2^{l_1 + l_2 + 1} q_{i,l_1}^+ q_{i,l_2}^+ \right) \label{equation:QUBO_model_2_1}
\end{split}
\end{align}
The second term of (\ref{equation:sylvester_final}) simplifies to:
\begin{align}\label{equation:QUBO_model_2_2}
    2 \sum_{i=1}^n \sum_{j=1}^n \sum_{k=1}^n  b_k a_{k,j} r_{j,i} y_i &\approx \sum_{i=1}^n \sum_{j=1}^n \sum_{k=1}^n b_k a_{k,j} r_{j,i} \left( -2^{m+2} q_i^- + \sum_{l=-m}^m 2^{l+1} q_{i,l}^+ \right)
\end{align}
Under this approximation technique, the QUBO model can be summarized as the summation of (\ref{equation:QUBO_model_2_1}) and (\ref{equation:QUBO_model_2_2}).
\begin{equation}
\begin{split}
    f(y) &= \sum_{i=1}^n d_{i,i} \left( 2^{2m+2} q_i^- + \sum_{l=-m}^m 2^{2l} q_{i,l}^+ - \sum_{l=-m}^m 2^{m+l+2} q_i^- q_{i,l}^+ + \sum_{l_1 < l_2} 2^{l_1 + l_2 + 1} q_{i,l_1}^+ q_{i,l_2}^+ \right) \\
    &+ \sum_{i=1}^n \sum_{j=1}^n \sum_{k=1}^n b_k a_{k,j} r_{j,i} \left( -2^{m+2} q_i^- + \sum_{l=-m}^m 2^{l+1} q_{i,l}^+ \right)
\end{split}
\end{equation}

After the solution $y^*$ for the $l^2$-norm minimization problem is obtained, we use the matrix $R$ to obtain the original solution to (\ref{equation:linear_system}):
\begin{equation}
    x^* = R y^*
\end{equation}

\subsection{Example: System of invertible linear equations in 2 variables} \label{sec:example_1}
Throughout the upcoming sections, we represent the variables of the linear system of equations using base-2 digits as given by O'Malley and Vesselinov (\ref{equation:QUBO_model_1}). We also assume that $A$, from the equation $Ax = b$, is an invertible matrix. 

We illustrate with an example to demonstrate the significance of applying Sylvester's law of inertia. This section closely follows the previous work by authors of this paper and their collaborators \cite{PLKWJ21}. Suppose we have a linear system $Ax = b$ given by \cite{JC21, PLKWJ21}:
\begin{equation} \label{equation:sample_example}
    \begin{pmatrix}
    3 & 1 \\
    -1 & 2
    \end{pmatrix}
    \begin{pmatrix}
    x_1 \\
    x_2
    \end{pmatrix}
    =
    \begin{pmatrix}
    -1 \\
    5
    \end{pmatrix}.
\end{equation}
The symmetric matrix $A^T A$ can be diagonalized to matrix $D$ using the invertible matrix $R$.
\begin{equation}
    D = \begin{pmatrix}
    \frac{8}{5} & 0 \\ 0 & \frac{98}{125}
    \end{pmatrix}, \; \; 
    R = \begin{pmatrix}
    \frac{2}{5} & -\frac{1}{25} \\
    0 & \frac{2}{5}
    \end{pmatrix}
\end{equation}
We use base-2 digits to approximate $y = (y_1, y_2)^T = R^{-1}x$:
\begin{equation} \label{equation:example_QAOA}
    y_i = \sum_{l=1}^3 2^{l-1} q_{i,l} - \sum_{l=1}^3 2^{l-1} q_{i,l+3}
\end{equation}
In accordance to (\ref{equation:qubit_product}), the set of qubits $\{q_{i,l}\}$ satisfy:
\begin{align} \label{equation:equivalence_relation}
\begin{cases}
    & q_{i,l_1} q_{i,l_2}= 0 \\
    & q_{i,j}^2 = q_{i,j}
\end{cases}
\end{align}
where $i \in \{1,2\}$ denotes the components of the variable $y$, $j \in \{1,2,\cdots,6\}$ the digits used in the base 2 representation of $y_i$, $1 \leq l_1 \leq 3$ the positive digits, and $4 \leq l_2 \leq 6$ the negative digits.

For a 2-dimensional system of linear equations, the energy level function from (\ref{equation:energy_function_simplify}) is given as
\begin{align} \label{equation:sylvester_final_1}
    f(y) = \sum_{i=1}^2 d_{i,i} y_i^2 - 2 \sum_{i=1}^2 \sum_{j=1}^2 \sum_{k=1}^2 b_k a_{k,j} r_{j,i} y_i,
\end{align}
where the column vector $y \in \mathbb{R}^n$ which satisfies
\begin{equation}
    f(y) = -b^T b = -26
\end{equation}
solves the desired system of linear equations. 
Substituting (\ref{equation:example_QAOA}) to (\ref{equation:sylvester_final_1}) and further reducing the terms using conditions on products of qubits from (\ref{equation:qubit_product}) and (\ref{equation:equivalence_relation}), we obtain:
%%%\textcolor{red}{Can we delete all squares?}
\begin{align}
\begin{split}
    f(y) = &8q_{11} + \frac{32}{5}q_{11}q_{12} + \frac{64}{5}q_{11}q_{13} + \frac{96}{5}q_{12} + \frac{128}{5}q_{12}q_{13} + \frac{256}{5}q_{13} - \frac{24}{5}q_{14} + \frac{32}{5}q_{14}q_{15} + \frac{64}{5}q_{14}q_{16} - \frac{32}{5}q_{15} \\
    &+  \frac{128}{5}q_{15}q_{16} - \frac{882}{125}q_{21} + \frac{392}{125}q_{21}q_{22} - \frac{1568}{125}q_{22} + \frac{784}{125}q_{21}q_{23} + \frac{1568}{125}q_{22}q_{23} - \frac{2352}{125}q_{23} + \frac{1078}{125}q_{24} \\
    &+ \frac{392}{125}q_{24}q_{25} + \frac{2352}{125}q_{25} + \frac{784}{125}q_{24}q_{26} + \frac{1568}{125}q_{25}q_{26} + \frac{5488}{125}q_{26}.
\end{split}
\end{align}
The above energy function gives rise to the $12 \times 12$ matrix $\hat{Q}$, defined as in (\ref{equation:Qhat_y}).
\begin{equation}\label{equation:Qhat_y}
{\small
    \hat{Q} =
    \begin{pmatrix}
    8 & 6.4 & 12.8  & 0  & 0 & 0 & 0 & 0 & 0 & 0 & 0 & 0 \\
    0  & 19.2  & 25.6  & 0  & 0 & 0 & 0 & 0 & 0 & 0 & 0 & 0 \\
    0  & 0  & 51.2 & 0 & 0 & 0 & 0 & 0 & 0 & 0 & 0 & 0 \\
    0  & 0  & 0  & -4.8 & 6.4 & 12.8 & 0 & 0 & 0 & 0 & 0 & 0 \\
    0  & 0  & 0  & 0  & -6.4 & 25.6 & 0 & 0 & 0 & 0 & 0 & 0 \\
    0  & 0  & 0  & 0  & 0  &  0 & 0 & 0 & 0 & 0 & 0 & 0 \\
    0  & 0  & 0  & 0  & 0  &  0 & -7.056 & 3.136 & 6.272 & 0 & 0 & 0 \\
    0  & 0  & 0  & 0  & 0  &  0 & 0 & -12.544 & 12.544 & 0 & 0 & 0 \\
    0  & 0  & 0  & 0  & 0  &  0 & 0 & 0 & -18.816 & 0 & 0 & 0 \\
    0  & 0  & 0  & 0  & 0  &  0 & 0 & 0 & 0 & 8.624 & 3.136 & 6.272 \\
    0  & 0  & 0  & 0  & 0  &  0 & 0 & 0 & 0 & 0 & 18.816 & 12.544 \\
    0  & 0  & 0  & 0  & 0  &  0 & 0 & 0 & 0 & 0 & 0 & 43.904 \\
    \end{pmatrix},
}
\end{equation}
The qubits utilized in the QUBO model are given by the column vector $q_y = [q_{11}, q_{12}, \cdots q_{26}]^T$, which approximates the unknown variables $y = (y_1, y_2)^T$ using base-2 digits. The cost function $f(y)$ can be reformulated in terms of $q_y$ as:
\begin{equation} \label{equation:new_qubo_model}
    f(y) = q_y^T \hat{Q} q_y = y^T D y - 2 b^T A R y.
\end{equation}
We note that the equality holds under a set of conditions specified in (\ref{equation:equivalence_relation}). Entanglements among pairs of qubits required for approximating $y_i$'s are represented as non-zero entries in the matrix $\hat{Q}$. Finding the vector of utilized qubits $q_y$ which satisfies $q_y^T \hat{Q} q_y = -26$ solves the given linear system.

We clearly observe that Sylvester's law of inertia and a number of conditions on qubit entanglements (\ref{equation:equivalence_relation}) forces a substantial portion of upper triangular matrix entries of $\hat{Q}$ to be equal to zero. This is unfortunately not the case for base-2 digit approximations of the variable $x$ from (\ref{equation:minimize_square}) \cite{JC21}. As before, we use $q_x = [q_{11},q_{12},\cdots,q_{26}]^T$ to represent the set of qubits utilized for approximating the variable $x = (x_1, x_2)^T$ using base-2 digits. The energy function formulating the system of lienar equations gives rise to the characterizing matrix $\hat{Q}'$:
\begin{equation}\label{equation:Qhat_prime_y}
{\small
    \hat{Q}' =
    \begin{pmatrix}
    26 & 40 & 80  & -20  & -40 & -80 & 2 & 4 & 8 & -2 & -4 & -8 \\
    0  & 72  & 160  & -40  & -80 & -160 & 4 & 8 & 16 & -4 & -8 & -16 \\
    0  & 0  & 224 & -80 & -160 & -320 & 8 & 16 & 32 & -8 & -16 & -32 \\
    0  & 0  & 0  & -6 & 40 & 80 & -2 & -4 & -8 & 2 & 4 & 8 \\
    0  & 0  & 0  & 0  & 8 & 160 & -4 & -8 & -16 & 4 & 8 & 16 \\
    0  & 0  & 0  & 0  & 0  &  96 & -8 & -16 & -32 & 8 & 16 & 32 \\
    0  & 0  & 0  & 0  & 0  &  0 & -13 & 20 & 40 & -10 & -20 & -40 \\
    0  & 0  & 0  & 0  & 0  &  0 & 0 & -16 & 80 & -20 & -40 & -80 \\
    0  & 0  & 0  & 0  & 0  &  0 & 0 & 0 & 8 & -40 & -80 & -160 \\
    0  & 0  & 0  & 0  & 0  &  0 & 0 & 0 & 0 & 23 & 20 & 40 \\
    0  & 0  & 0  & 0  & 0  &  0 & 0 & 0 & 0 & 0 & 56 & 80 \\
    0  & 0  & 0  & 0  & 0  &  0 & 0 & 0 & 0 & 0 & 0 & 152 \\
    \end{pmatrix},
}
\end{equation}
where $x^T \hat{Q}' x = x^T A^T A x - 2 b^T A x$ up equivalence relation $q_{i,j}^2 = q_{i,j}$. Then, we have 
\begin{equation} \label{equation:vanilla_qubo_model}
    q_x^T \hat{Q}' q_x = x^T A^T A x - 2 b^T A x
\end{equation}
The qubit entanglements required for representing $x_i$'s correspond to upper triangular entries of the matrix $\hat{Q}'$. The solution to the quadratic equation $q_x^T \hat{Q}' q_x = -26$ is the solution to the linear system (\ref{equation:sample_example}).

\section{Implementation}

This section verifies how Sylvester's law of inertia significantly boosts both effectiveness and extensiveness of QUBO formulations of systems of linear equations on a quantum annealing device. For all experiments we use the D-Wave 2000Q quantum annealer to process the system of linear equations. We approximate the unknown variables of a given system of linear equations using O'Malley and Vesselinov's technique \cite{MV16} to effectively compare the proposed QUBO formulation to previously studied models.

\subsection{Effectiveness} \label{sec: implementation_effectiveness}

As shown in previous research \cite{PLKWJ21}, we process the system of linear equations in 2 variables from Section \ref{sec:example_1} on the D-Wave 2000Q quantum annealer. We perform 3 trials of both QUBO models (\ref{equation:new_qubo_model}, \ref{equation:vanilla_qubo_model}) on the D-Wave system using 10,000 anneals to verify the strength of utilizing Sylvester's theorem. 

\subsubsection{Vanilla model}
The vanilla QUBO model (\ref{equation:vanilla_qubo_model}) computes all possible combinations of qubits $q_x$ for $(x_1, x_2) = (-1, 2)$, which are solutions obtained from the 2-dimensional linear system without using Sylvester's law of inertia. There are 7 qubit combinations for $x_1 = -1$, and 6 qubit combinations for $x_2 = 2$ under the base-2 representation from (\ref{equation:example_QAOA}).
\begin{align} \label{equation:qubits_all_combinations}
    \begin{split}
        (q_{11}, q_{12}, q_{13}, q_{14}, q_{15}, q_{16}) \in \{ &(0,0,0,1,0,0),(0,1,0,1,1,0), \\ 
        &(0,0,1,1,0,1),(0,1,1,1,1,1), \\
        &(1,0,0,0,1,0),(1,0,1,0,1,1), \\
        &(1,1,0,0,0,1) \} \\
        (q_{21},q_{22},q_{23},q_{24},q_{25},q_{26}) \in \{ &(0,0,1,0,1,0),(0,1,0,0,0,0),\\
        &(0,1,1,0,0,1),(1,0,1,1,1,0),\\
        &(1,1,0,1,0,0),(1,1,1,1,0,1) \}
    \end{split}
\end{align}
The implementation results of the vanilla model is provided in Table \ref{tab:original_QUBO}. We abbreviate the set of all possible qubit combinations of $x_2 = 2$. Each row lists the number of anneals which achieves the given combination of qubits and minimizes the energy function. The vanilla model minimizes the energy function 887, 1181, and 1065 times out of 10,000 anneals.

\begin{table}[]
\begin{center}
{\small
\begin{tabular}{c|c|c|c|c|c||c|c|c|c|c|c||c||c|c|c}
\hline
\hline
\multirow{2}{*}{$q_{11}$} & \multirow{2}{*}{$q_{12}$} & \multirow{2}{*}{$q_{13}$} & \multirow{2}{*}{$q_{14}$} & \multirow{2}{*}{$q_{15}$} & \multirow{2}{*}{$q_{16}$} & \multirow{2}{*}{$q_{21}$} & \multirow{2}{*}{$q_{22}$} & \multirow{2}{*}{$q_{23}$} & \multirow{2}{*}{$q_{24}$} & \multirow{2}{*}{$q_{25}$} & \multirow{2}{*}{$q_{26}$} & \multirow{2}{*}{Energy} & \multicolumn{3}{c}{\# Occurrences} \\ \cline{14-16} & & & & & & & & & & & & & Run 1 & Run 2 & Run 3 \\ \hline \hline
0 & 0 & 0 & 1 & 0 & 0 & \multicolumn{6}{c||}{All 6 combinations} & -26.0 & 203 & 66 & 50 \\ \hline
0 & 1 & 0 & 1 & 1 & 0 & \multicolumn{6}{c||}{All 6 combinations} & -26.0 & 77 & 49 & 531 \\ \hline
0 & 0 & 1 & 1 & 0 & 1 & \multicolumn{6}{c||}{All 6 combinations} & -26.0 & 131 & 147 & 251 \\ \hline 
0 & 1 & 1 & 1 & 1 & 1 & \multicolumn{6}{c||}{All 6 combinations} & -26.0 & 71 & 116 & 51 \\ \hline
1 & 0 & 0 & 0 & 1 & 0 & \multicolumn{6}{c||}{All 6 combinations} & -26.0 & 75 & 43 & 74 \\ \hline 
1 & 0 & 1 & 0 & 1 & 1 & \multicolumn{6}{c||}{All 6 combinations} & -26.0 & 71 & 83 & 62 \\ \hline
1 & 1 & 0 & 0 & 0 & 1 & \multicolumn{6}{c||}{All 6 combinations} & -26.0 & 259 & 677 & 46 \\ \hline
\hline 
\multicolumn{11}{c}{} & & Total & 887 & 1181 & 1065 \\ \hline
\hline
\end{tabular}
}
\end{center}
\caption{Number of occurrences which minimizes the energy function obtained from (\ref{equation:Qhat_prime_y}) \cite{PLKWJ21}. The list of all possible qubit combinations for $x_2 = 2$ is omitted from the table.}
\label{tab:original_QUBO}
\end{table}

\subsubsection{Proposed model}
Sylvester's law of inertia transforms the solution of the linear system of equations (\ref{equation:sample_example}) to $(y_1, y_2) = (-2,5)$. The objective of the new simplified QUBO model (\ref{equation:new_qubo_model}) is to obtain the unique combination of qubits $y_i = q_{i1} + 2q_{i2} + 4q_{i3} - q_{i4} - 2q_{i5} - 4q_{i6}$:
\begin{align}
    \begin{split}
        (q_{11}, q_{12}, q_{13}, q_{14}, q_{15}, q_{16}) &= (0, 0, 0, 0, 1, 0) \\
        (q_{21}, q_{22}, q_{23}, q_{24}, q_{25}, q_{26}) &= (1, 0, 1, 0, 0, 0) 
    \end{split}
\end{align}
Specifying the zero terms of the matrix characterizing the system of linear equations may affect the performance of the quantum algorithm on the D-wave computer. To verify this, the proposed model can be executed in two ways. One way is to explicitly code the zero terms appearing on the upper triangular region of the characterizing matrix $\hat{Q}$ to the D-wave system; the other way is to omit them. We display the number of occurrences of the desired qubits in Table \ref{tab:new_QUBO_with_0}. The former minimizes the energy function 1526, 2495, and 2063 times out of 10,000 anneals. On the other hand, the latter obtains the lowest energy level for 2103, 4441, and 1727 anneals out of 10,000 anneals.

\begin{table}[]
\centering
{\small
\begin{tabular}{c|c|c|c|c|c||c|c|c|c|c|c||c|c||c|c|c}
\hline
\hline
\multirow{2}{*}{$q_{11}$} & \multirow{2}{*}{$q_{12}$} & \multirow{2}{*}{$q_{13}$} & \multirow{2}{*}{$q_{14}$} & \multirow{2}{*}{$q_{15}$} & \multirow{2}{*}{$q_{16}$} & \multirow{2}{*}{$q_{21}$} & \multirow{2}{*}{$q_{22}$} & \multirow{2}{*}{$q_{23}$} & \multirow{2}{*}{$q_{24}$} & \multirow{2}{*}{$q_{25}$} & \multirow{2}{*}{$q_{26}$} & \multirow{2}{*}{Energy} & \multirow{2}{*}{Zero Terms} & \multicolumn{3}{c}{\# Occurrences} \\ \cline{15-17} & & & & & & & & & & & & & & Run 1 & Run 2 & Run 3 \\ \hline \hline
0 & 0 & 0 & 0 & 1 & 0 & 1 & 0 & 1 & 0 & 0 & 0 & -26.0 & Yes & 1526 & 2495 & 2063 \\ \hline
0 & 0 & 0 & 0 & 1 & 0 & 1 & 0 & 1 & 0 & 0 & 0 & -26.0 & No & 2103 & 4441 & 1727 \\ \hline
\hline 
\end{tabular}
}
\centering
\caption{Number of occurrences which minimizes the energy function obtained from (\ref{equation:Qhat_y}) \cite{PLKWJ21}. The column ``Zero Terms" specifies whether the zero terms of the characterizing matrix are explicitly implemented on the D-wave system.}
\label{tab:new_QUBO_with_0}
\end{table}

\subsection{Extensiveness} \label{sec: implementation_extensiveness}

We shall now empirically assess the comparative extensiveness of the proposed QUBO model (\ref{equation:energy_function_simplify}) to that of the vanilla QUBO model (\ref{equation:energy_function_defn}) in processing a system of linear equations. Consider the following $4$ systems of linear equations in $2,3,4,$ and $5$ variables:
\begin{equation} \label{equation:2_variables}
    \begin{pmatrix}
    3 & 1 \\
    -1 & 2
    \end{pmatrix}
    \begin{pmatrix}
    x_1 \\
    x_2
    \end{pmatrix}
    =
    \begin{pmatrix}
    -1 \\
    5
    \end{pmatrix}
\end{equation}

\begin{equation} \label{equation:3_variables}
    \begin{pmatrix}
    5 & 0 & 1 \\
    -1 & 2 & 1 \\
    3 & 2 & 3
    \end{pmatrix}
    \begin{pmatrix}
    x_1 \\
    x_2 \\
    x_3
    \end{pmatrix}
    =
    \begin{pmatrix}
    -65 \\
    -15 \\
    -80
    \end{pmatrix}
\end{equation}

\begin{equation} \label{equation:4_variables}
    \begin{pmatrix}
   5 & 0 & 1 & 3 \\
   -1 & 2 & 1 & 4 \\
   3 & 2 & 3 & 5 \\
   6 & 29 & 30 & -14
   \end{pmatrix}
   \begin{pmatrix}
   x_1 \\
   x_2 \\
   x_3 \\
   x_4
   \end{pmatrix}
   =
   \begin{pmatrix}
   -65 \\
   -15 \\
   -80 \\
   -1
   \end{pmatrix}
\end{equation}

\begin{equation} \label{equation:5_variables}
    \begin{pmatrix}
    5 & 0 & 1 & 3 & 8 \\
   -1 & 2 & 1 & 4 & 5 \\
   3 & 2 & 3 & 5 & 26 \\
   6 & 29 & 30 & -14 & -1 \\
   5 & 3 & 4 & -2 & 65
    \end{pmatrix}
    \begin{pmatrix}
    x_1 \\
    x_2 \\
    x_3 \\
    x_4 \\
    x_5
    \end{pmatrix}
    =
    \begin{pmatrix}
    -65 \\
    -15 \\
    -80 \\
    -1 \\
    47
    \end{pmatrix}
\end{equation}
We determine the extensiveness of both QUBO models (\ref{equation:energy_function_defn}, \ref{equation:energy_function_simplify}) by verifying whether the D-wave quantum annealer using $1,000$ anneals compiles the given system of linear equations (\ref{equation:2_variables}, \ref{equation:3_variables}, \ref{equation:4_variables}, \ref{equation:5_variables}) with predetermined number of base-2 digits representing the unknown variables. For example, a system of linear equation in $2$ variables where each variable is approximated with $20$ base-2 digits requires the QUBO model to utilize $20 \times 2 = 40$ qubits on the D-wave quantum annealer. Exemplary compilations of these systems of linear equations can be found in the following github repository: https://github.com/ktfriends/Quantum\_Computing/tree/main/Sylvester\_Theorem

\subsubsection{Vanilla model}

As shown in Table \ref{tab:vanilla_extensiveness}, the D-wave quantum annealer compiles the vanilla QUBO model (\ref{equation:energy_function_defn}) which utilizes at most $64$ qubits. We omitted the experimental results obtained for QUBO models utilizing more than $80$ qubits, for all these attempts failed to compile on the D-wave quantum annealer.

\begin{table}[]
    \centering
    \begin{tabular}{c|c|c|c||c}
        Linear system & \# variables & \# base-2 digits & \# utilized qubits & Compile \\
        \hline
        \hline
        (\ref{equation:2_variables}) & 2 & 20 & 40 & $\circ$ \\
        \hline
        (\ref{equation:3_variables}) & 3 & 20 & 60 & $\circ$ \\
        \hline
        (\ref{equation:4_variables}) & 4 & 20 & 80 & $\times$ \\
        \hline
        \hline
        (\ref{equation:2_variables}) & 2 & 6 & 12 & $\circ$ \\
        \hline
        (\ref{equation:2_variables}) & 2 & 32 & \textbf{64} & $\circ$ \\
        \hline
        (\ref{equation:2_variables}) & 2 & 34 & 68 & $\times$ \\
        \hline
        \hline
    \end{tabular}
    \caption{Number of qubits implementable on a D-wave 2000Q quantum annealer using the vanilla QUBO model (\ref{equation:energy_function_defn}). Experimental results suggest that the vanilla QUBO model can utilize at most $64$ qubits.}
    \label{tab:vanilla_extensiveness}
\end{table}

\subsubsection{Proposed model}

Experimental results suggest that the D-wave quantum annealer compiles the proposed QUBO model (\ref{equation:energy_function_simplify}) utilizing at most $132$ qubits, as demonstrated in Table \ref{tab:new_extensiveness}.

\begin{table}[]
    \centering
    \begin{tabular}{c|c|c|c||c}
        Linear system & \# variables & \# base-2 digits & \# utilized qubits & Compile \\
        \hline
        \hline
        (\ref{equation:2_variables}) & 2 & 20 & 40 & $\circ$ \\
        \hline
        (\ref{equation:3_variables}) & 3 & 20 & 60 & $\circ$ \\
        \hline
        (\ref{equation:4_variables}) & 4 & 20 & 80 & $\circ$ \\
        \hline
        (\ref{equation:5_variables}) & 5 & 20 & 100 & $\circ$ \\
        \hline
        \hline
        (\ref{equation:2_variables}) & 2 & 6 & 12 & $\circ$ \\
        \hline
        (\ref{equation:2_variables}) & 2 & 60 & 120 & $\circ$ \\
        \hline
        (\ref{equation:2_variables}) & 2 & 66 & \textbf{132} & $\circ$ \\
        \hline
        (\ref{equation:2_variables}) & 2 & 68 & 136 & $\times$ \\
        \hline
        (\ref{equation:2_variables}) & 2 & 70 & 140 & $\times$ \\
        \hline
        (\ref{equation:2_variables}) & 2 & 80 & 160 & $\times$ \\
        \hline
        (\ref{equation:2_variables}) & 2 & 100 & 200 & $\times$ \\
        \hline
        \hline
    \end{tabular}
    \caption{Number of qubits implementable on a D-wave 2000Q quantum annealer using the proposed QUBO model (\ref{equation:energy_function_simplify}). Experimental results suggest that the proposed QUBO model can utilize at most $132$ qubits.}
    \label{tab:new_extensiveness}
\end{table}

\section{Discussion}
To examine whether the proposed model is effective for processing high dimensional systems of linear equations, the experimental results should verify the following key criteria:
\begin{itemize}
    \item \textbf{Effectiveness}: Guarantees enhanced and cost-efficient performance with the simplified QUBO model
    \item \textbf{Extensiveness}: Extends implementable QUBO model with reduced qubit connectivity
\end{itemize}
In this section, we carefully examine and assess how the implementation results from the previous section achieve all three criteria for extending the number of implementable variables of systems of linear equations. We expect that the proposed model effectively enhances efficiency, accuracy, and extensiveness of previously studied QUBO models for solving a system of linear equations on a quantum annealing device. 

\subsection{Effectiveness and the simplified characterization matrix}

As demonstrated in Section \ref{sec: implementation_effectiveness}, both implementations of the proposed QUBO model (\ref{equation:new_qubo_model}) clearly surpasses the vanilla QUBO model (\ref{equation:vanilla_qubo_model}) \cite{PLKWJ21}. A list of average number of occurrences and the average probability of obtaining the lowest energy levels obtained from all three experiments is provided in Table \ref{tab:Results Comparison}.

\begin{table*}[hbt!]
\centering
{\small
\begin{tabular}{c||c|c|c}
\hline
\hline
\multirow{2}{*}{\# Trial} & Vanilla QUBO model & New QUBO model & New QUBO model \\ 
& (Table \ref{tab:original_QUBO}) & (Table \ref{tab:new_QUBO_with_0}, row 1) & (Table \ref{tab:new_QUBO_with_0}, row 2) \\ \hline
\hline
Run 1 & 887 & 1526 & 2103 \\ \hline
Run 2 & 1181 & 2495 & 4441 \\ \hline
Run 3 & 1065 & 2063 & 1727 \\ \hline
\hline
Average \# Occurrences & 1044 & 2028 & 2758 \\ \hline
Average Probability & 10.44\% & 20.28 \% & 27.58 \% \\
\hline
\hline
\end{tabular}
}
\centering
\caption{A summary of the average number of occurrences that minimize the energy level function from Section \ref{sec: implementation_effectiveness} for all three models \cite{PLKWJ21}.}
\label{tab:Results Comparison}
\end{table*}

The new QUBO model minimizes the energy function which formulates (\ref{equation:sample_example}) with average probability ranging between $20.28 \%$ and $27.58\%$. These values are at least twice to that obtained from the vanilla QUBO model, which is $10.44 \%$. Among the two implementations of the new QUBO model, omitting the zero entries from the code significantly improves the accuracy of the algorithm.

The block diagonalization of the matrix $Q$ characterizing the QUBO model is a key contributor to the outperformance of the new QUBO model. Suppose we want to obtain a solution of a system of $n$ linear equations with $n$ variables. We further assume that we use $2m$ digits to represent the variables using base-2 digits (\ref{equation:QUBO_model_1}). The matrix $Q$ characterizing the QUBO model of the systems of linear equations is a $2mn \times 2mn$-matrix.

The number of non-zero entries of $Q_{vanilla}$ characterizing the vanilla QUBO model, as in (\ref{equation:vanilla_qubo_model}), satisfies the following inequality:
\begin{equation}
    \# \text{non-zero entries of } Q_{vanilla} \leq \frac{2mn(2mn+1)}{2} = mn(2mn+1).
\end{equation}
Sylvester's law of inertia allows $Q_{vanilla}$ to be decomposed into $2n$ block matrices of dimension $m \times m$, all of which are upper triangular.The number of non-zero entries of the newly obtained matrix $Q_{new}$, as in (\ref{equation:new_qubo_model}), satisfies
\begin{equation}
    \# \text{non-zero entries of } Q_{new} \leq 2n \times \frac{m(m+1)}{2} = mn(m+1).
\end{equation}
Therefore, assuming that both matrices $Q_{vanilla}$ and $Q_{new}$ have maximal numbers of non-zero entries, Sylvester's law of inertia significantly reduces the number of non-zero entries by at least a factor of $\frac{1}{2n-1}$, and at most a factor of $\frac{1}{2n}$:
\begin{equation}
    \frac{1}{2n} \leq \frac{\# \text{non-zero entries of } Q_{new}}{\# \text{non-zero entries of } Q_{vanilla}} = \frac{m+1}{2mn+1} \leq \frac{1}{2n-1}.
\end{equation}
For instance, when $n=2$ and $m=3$, there are at most $78$ non-zero entries in $Q_{vanilla} = \hat{Q}'$ from (\ref{equation:Qhat_prime_y}). However, there are at most $24$ non-zero entries in $Q_{new} = \hat{Q}$ from (\ref{equation:Qhat_prime_y}), the number of which is less than a third of 78.

\subsection{Extensiveness and reduced qubit connectivity}

The strength of the proposed QUBO model can be also found from its extensiveness in compiling a wide variety of systems of linear equations. Further implementations of linear systems on D-wave quantum annealer, as demonstrated in Section \ref{sec: implementation_extensiveness}, clearly show that the proposed model allows compilations of higher dimensional systems of linear equations that previous models were not able to achieve. In fact, experimental verification suggests that the proposed model doubles the utilizable number of qubits compared to other QUBO formulations.

Such expansions in a range of implementable system of equations originates from substantial reduction in the connectivity of qubits required to perform the QUBO model on quantum computers. The key philosophy behind the proposed methodology is to transform a dense matrix characterizing the systems of linear equations to a sparse matrix using matrix congruence relations. This allows room for implementing a wider range of systems of linear equations on a quantum computer without substantially increasing the number of utilized qubits.

\begin{figure}
    \centering
    \resizebox{\columnwidth}{!}{
    \begin{tikzpicture}
 \graph [simple] { 
      subgraph K_n [n=12, clockwise, radius=3cm, nodes={draw, inner sep=1pt, fill=white, circle, minimum size=0.5cm}];
    };
    \end{tikzpicture}
    \hspace{1cm}
    \begin{tikzpicture}
 \graph [simple] { 
      subgraph K_n [n=12, clockwise, radius=3cm, nodes={draw, inner sep=1pt, fill=white, circle, minimum size=0.5cm}];
      1 -!- {4,5,6,7,8,9,10,11};
      2 -!- {4,5,6,7,8,9,10,11,12};
      3 -!- {5,6,7,8,9,10,11,12};
      4 -!- {1,2,7,8,9,10,11,12};
      5 -!- {1,2,3,7,8,9,10,11,12};
      6 -!- {1,2,3,8,9,10,11,12};
      7 -!- {1,2,3,4,5,10,11,12};
      8 -!- {1,2,3,4,5,6,10,11,12};
      9 -!- {1,2,3,4,5,6,11,12};
      10 -!- {1,2,3,4,5,6,7,8};
      11 -!- {1,2,3,4,5,6,7,8,9};
      12 -!- {2,3,4,5,6,7,8,9};
    };
    \end{tikzpicture}
    }
    \caption{A graph of connectivity of $12$ qubits required to simultaneously process the characterizing matrix $\hat{Q}'$ of the vanilla QUBO model (left) and $\hat{Q}$ of the proposed QUBO model (right). The QUBO model formulates a system of linear equations in $2$ variables where each variable is approximated by $6$ base-2 digits. Note that $3$ out of the $6$ qubits are used to represent positive digits, whereas the other $3$ are used to represent negative digits.}
    \label{graph:qubit_connectivity_12}
\end{figure}
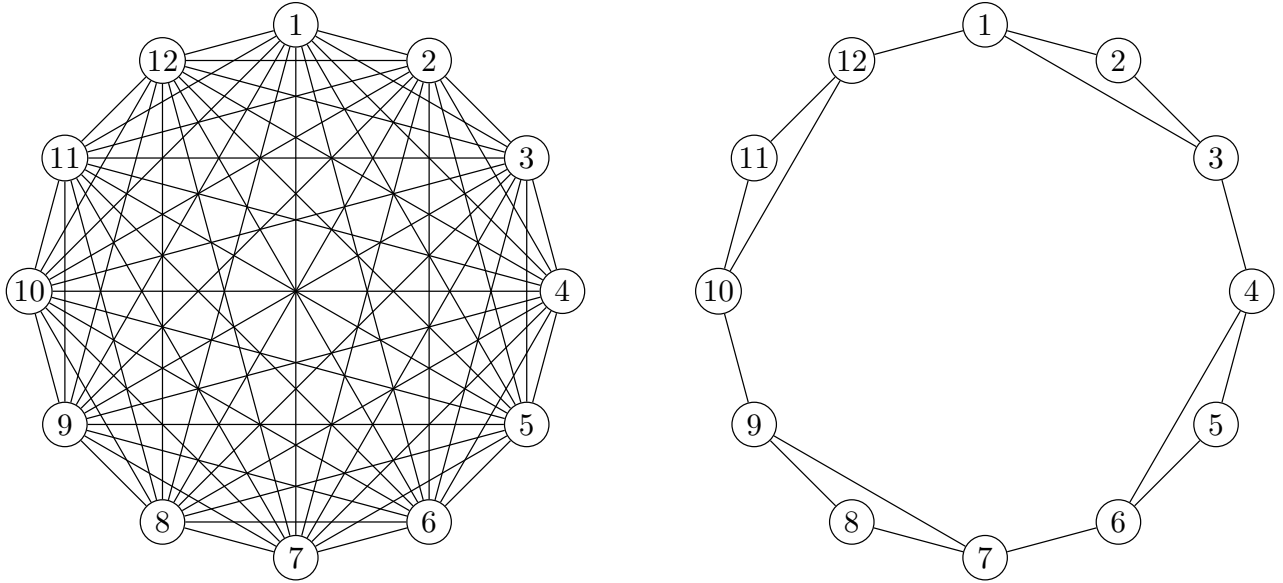

We demonstrate the aformentioned advantage with the characterizing matrices $\hat{Q}$ and $\hat{Q}'$ from (\ref{equation:Qhat_prime_y}) and (\ref{equation:Qhat_y}). For now, let's assume that all qubit quadratic terms appearing in the energy function (\ref{equation:new_qubo_model}) let the corresponding pair of qubits to be simultaneously entangled. The matrix $\hat{Q}'$ from the vanilla QUBO model (\ref{equation:Qhat_prime_y}), requires simultaneous couplings of all pairs of 12 qubits, as shown in the graph on the left of Figure \ref{graph:qubit_connectivity_12} (We note that the graphs are drawn in a similar fashion to those appearing on a previous work by Zbinden et al. \cite{ZBDE20}).

\begin{figure}
    \centering
    \resizebox{\columnwidth}{!}{
    \begin{tikzpicture}
 \graph [simple] { 
      subgraph K_n [n=40, clockwise, radius=10cm, nodes={draw, inner sep=1pt, fill=white, circle, minimum size=1cm}];
    };
    \end{tikzpicture}
    \hspace{1cm}
    \begin{tikzpicture}
 \graph [simple] { 
      subgraph K_n [n=40, clockwise, radius=10cm, nodes={draw, inner sep=1pt, fill=white, circle, minimum size=1cm}];
      1 -!- {11,12,13,14,15,16,17,18,19,20,21,22,23,24,25,26,27,28,29,30,31,32,33,34,35,36,37,38,39};
      2 -!- {11,12,13,14,15,16,17,18,19,20,21,22,23,24,25,26,27,28,29,30,31,32,33,34,35,36,37,38,39,40};
      3 -!- {11,12,13,14,15,16,17,18,19,20,21,22,23,24,25,26,27,28,29,30,31,32,33,34,35,36,37,38,39,40};
      4 -!- {11,12,13,14,15,16,17,18,19,20,21,22,23,24,25,26,27,28,29,30,31,32,33,34,35,36,37,38,39,40};
      5 -!- {11,12,13,14,15,16,17,18,19,20,21,22,23,24,25,26,27,28,29,30,31,32,33,34,35,36,37,38,39,40};
      6 -!- {11,12,13,14,15,16,17,18,19,20,21,22,23,24,25,26,27,28,29,30,31,32,33,34,35,36,37,38,39,40};
      7 -!- {11,12,13,14,15,16,17,18,19,20,21,22,23,24,25,26,27,28,29,30,31,32,33,34,35,36,37,38,39,40};
      8 -!- {11,12,13,14,15,16,17,18,19,20,21,22,23,24,25,26,27,28,29,30,31,32,33,34,35,36,37,38,39,40};
      9 -!- {11,12,13,14,15,16,17,18,19,20,21,22,23,24,25,26,27,28,29,30,31,32,33,34,35,36,37,38,39,40};
      10 -!- {12,13,14,15,16,17,18,19,20,21,22,23,24,25,26,27,28,29,30,31,32,33,34,35,36,37,38,39,40};
      11 -!- {1,2,3,4,5,6,7,8,9,21,22,23,24,25,26,27,28,29,30,31,32,33,34,35,36,37,38,39,40};
      12 -!- {1,2,3,4,5,6,7,8,9,10, 21,22,23,24,25,26,27,28,29,30,31,32,33,34,35,36,37,38,39,40};
      13 -!- {1,2,3,4,5,6,7,8,9,10, 21,22,23,24,25,26,27,28,29,30,31,32,33,34,35,36,37,38,39,40};
      14 -!- {1,2,3,4,5,6,7,8,9,10, 21,22,23,24,25,26,27,28,29,30,31,32,33,34,35,36,37,38,39,40};
      15 -!- {1,2,3,4,5,6,7,8,9,10, 21,22,23,24,25,26,27,28,29,30,31,32,33,34,35,36,37,38,39,40};
      16 -!- {1,2,3,4,5,6,7,8,9,10, 21,22,23,24,25,26,27,28,29,30,31,32,33,34,35,36,37,38,39,40};
      17 -!- {1,2,3,4,5,6,7,8,9,10, 21,22,23,24,25,26,27,28,29,30,31,32,33,34,35,36,37,38,39,40};
      18 -!- {1,2,3,4,5,6,7,8,9,10, 21,22,23,24,25,26,27,28,29,30,31,32,33,34,35,36,37,38,39,40};
      19 -!- {1,2,3,4,5,6,7,8,9,10, 21,22,23,24,25,26,27,28,29,30,31,32,33,34,35,36,37,38,39,40};
      20 -!- {1,2,3,4,5,6,7,8,9,10,22,23,24,25,26,27,28,29,30,31,32,33,34,35,36,37,38,39,40};
      21 -!- {1,2,3,4,5,6,7,8,9,10,11,12,13,14,15,16,17,18,19, 31,32,33,34,35,36,37,38,39,40};
      22 -!- {1,2,3,4,5,6,7,8,9,10,11,12,13,14,15,16,17,18,19,20, 31,32,33,34,35,36,37,38,39,40};
      23 -!- {1,2,3,4,5,6,7,8,9,10,11,12,13,14,15,16,17,18,19,20, 31,32,33,34,35,36,37,38,39,40};
      24 -!- {1,2,3,4,5,6,7,8,9,10,11,12,13,14,15,16,17,18,19,20, 31,32,33,34,35,36,37,38,39,40};
      25 -!- {1,2,3,4,5,6,7,8,9,10,11,12,13,14,15,16,17,18,19,20, 31,32,33,34,35,36,37,38,39,40};
      26 -!- {1,2,3,4,5,6,7,8,9,10,11,12,13,14,15,16,17,18,19,20, 31,32,33,34,35,36,37,38,39,40};
      27 -!- {1,2,3,4,5,6,7,8,9,10,11,12,13,14,15,16,17,18,19,20, 31,32,33,34,35,36,37,38,39,40};
      28 -!- {1,2,3,4,5,6,7,8,9,10,11,12,13,14,15,16,17,18,19,20, 31,32,33,34,35,36,37,38,39,40};
      29 -!- {1,2,3,4,5,6,7,8,9,10,11,12,13,14,15,16,17,18,19,20, 31,32,33,34,35,36,37,38,39,40};
      30 -!- {1,2,3,4,5,6,7,8,9,10,11,12,13,14,15,16,17,18,19,20,32,33,34,35,36,37,38,39,40};
      31 -!- {1,2,3,4,5,6,7,8,9,10,11,12,13,14,15,16,17,18,19,20,21,22,23,24,25,26,27,28,29};
      32 -!- {1,2,3,4,5,6,7,8,9,10,11,12,13,14,15,16,17,18,19,20,21,22,23,24,25,26,27,28,29,30 };
      33 -!- {1,2,3,4,5,6,7,8,9,10,11,12,13,14,15,16,17,18,19,20,21,22,23,24,25,26,27,28,29,30 };
      34 -!- {1,2,3,4,5,6,7,8,9,10,11,12,13,14,15,16,17,18,19,20,21,22,23,24,25,26,27,28,29,30 };
      35 -!- {1,2,3,4,5,6,7,8,9,10,11,12,13,14,15,16,17,18,19,20,21,22,23,24,25,26,27,28,29,30 };
      36 -!- {1,2,3,4,5,6,7,8,9,10,11,12,13,14,15,16,17,18,19,20,21,22,23,24,25,26,27,28,29,30 };
      37 -!- {1,2,3,4,5,6,7,8,9,10,11,12,13,14,15,16,17,18,19,20,21,22,23,24,25,26,27,28,29,30 };
      38 -!- {1,2,3,4,5,6,7,8,9,10,11,12,13,14,15,16,17,18,19,20,21,22,23,24,25,26,27,28,29,30 };
      39 -!- {1,2,3,4,5,6,7,8,9,10,11,12,13,14,15,16,17,18,19,20,21,22,23,24,25,26,27,28,29,30 };
      40 -!- {2,3,4,5,6,7,8,9,10,11,12,13,14,15,16,17,18,19,20,21,22,23,24,25,26,27,28,29,30 };
    };
    \end{tikzpicture}
    }
    \caption{A graph of connectivity of $40$ qubits required to simultaneously process the characterizing matrix of the vanilla QUBO model (left) and that of the proposed QUBO model (right). The QUBO model formulates a system of linear equations in $2$ variables where each variable is approximated by $20$ base-2 digits. Note that $10$ out of the $20$ qubits are used to represent positive digits, whereas the other $10$ are used to represent negative digits.}
    \label{graph:qubit_connectivity_40}
\end{figure}
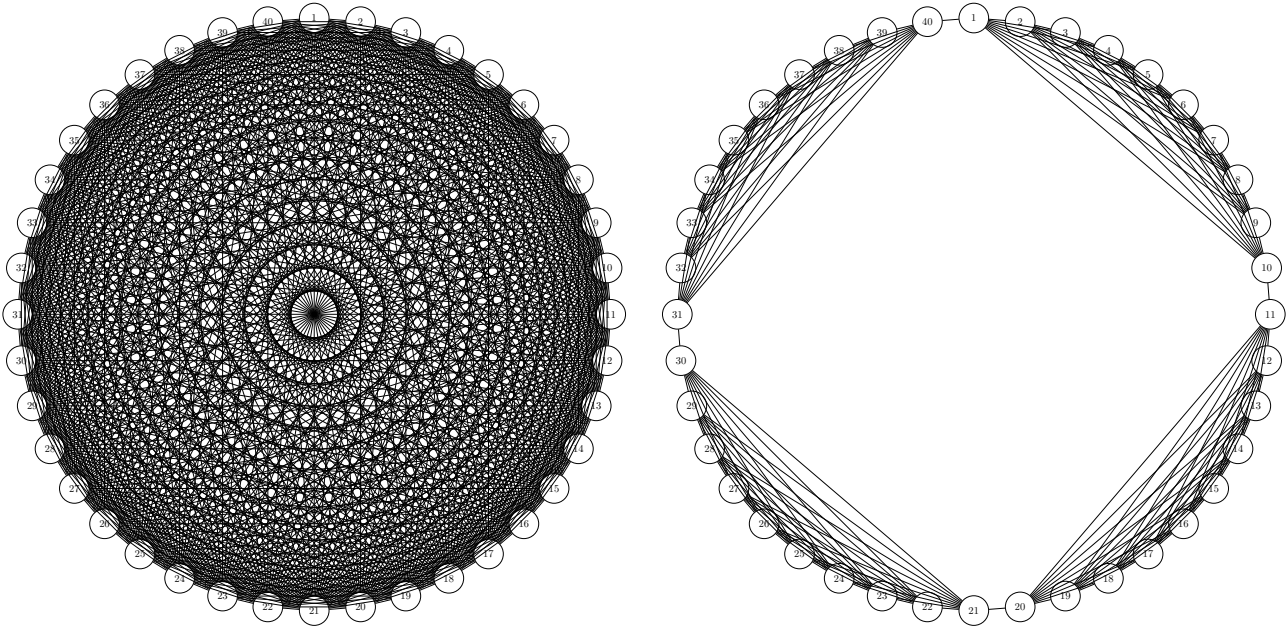

On the other hand, our proposed model is characterized by the matrix $\hat{Q}$ from (\ref{equation:Qhat_y}), which can be decomposed into 4 blocks of $3 \times 3$ upper triangular matrices. This implies that any two qubits which process different blocks of matrices do not have to be entangled. Therefore, we require that 3 out of 12 qubits have to be interconnected on the quantum computer. For example, the following diagram of qubit connections is sufficient to simultaneously process the matrix $\hat{Q}$. We see that only a set of three qubits ($\{1,2,3\}, \{4,5,6\}, \{7,8,9\}, \{10,11,12\}$) are fully connected. The diagram of qubit connections displayed in the graph on the right of Figure \ref{graph:qubit_connectivity_12} is far simpler than the one required for implementing the vanilla QUBO model. The stark difference between the connectivity of qubits required for two QUBO formulations becomes clearly noticeable for system of linear equations utilizing $40$ qubits, see Figure \ref{graph:qubit_connectivity_40} for instance.

In general, suppose that we want to solve a system of $n$ linear equations with $n$ variables. We further assume that we use $m$ positive digits and $m$ negative digits to approximate unknown variables using base 2 digits. The matrix $Q_{vanilla}$ characterizing the vanilla QUBO model is a $2mn \times 2mn$ upper triangular matrix. Therefore, a quantum computer would require a fully connected network of $2mn$ qubits to simultaneously process $Q_{vanilla}$. On the other hand, the matrix $Q_{new}$ characterizing the proposed QUBO model can be decomposed into $2n$ upper triangular block matrices of dimension $m \times m$. This implies that a quantum computer would only require $2n$ sets of fully connected network of $2m$ qubits to simultaneously process $Q_{vanilla}$. In other words, our proposed method can reformulate the QUBO model in accordance to the given qubit connectivity of commercial quantum computers. 

\begin{table}[]
    \centering
    \begin{tabular}{c|c}
        QUBO model & Qubit Connectivity \\
        \hline
        \hline
        Vanilla & Fully connected network of $2mn$ qubits \\
        \hline
        Proposed & $2n$ sets of fully connected network of $m$ qubits \\
        \hline
    \end{tabular}
    \caption{A table summarizing the required qubit connectivity for simulatenously processing the characterizing matrix of the QUBO model for a system of linear equations with $n$ variables}
    \label{tab:qubit_connectivity}
\end{table}

The underlying topology of connectivity of qubits on the D-wave 2000Q quantum annealer is a key factor that determines the upper bound on the number of qubits the given QUBO formulation can execute. The D-wave 2000Q quantum annealer uses the Chimera topology $C_{16,16,4}$ to model the connectivity of $2048$ qubits \cite{BKR16, BKR21}. Given a positive number $m$, the space $C_{m,m,4}$, comprised of $4m^2$ qubits, is an interconnected $m \times m$ grid of bipartite graphs $K_{4,4}$. There exists a natural embedding of the complete graph with $4m$ nodes $K_{4m}$ inside $C_{m,m,4}$ as a graph minor, corresponding to a triangle on the $m \times m$ grid consisting of $4m$ chains of length $m+1$ \cite{BKR16, C11}. As for the topology $C_{16,16,4}$, the space has a naturally embedded complete graph $K_{64}$ inside $C_{16,16,4}$ as a graph minor.
\begin{equation}
    K_{64} \hookrightarrow C_{16,16,4}
\end{equation}

Observant readers may have already noticed that the number of nodes of the complete graph that can be embedded in $C_{16,16,4}$ is precisely the empirical maximum number of qubits the vanilla QUBO model can process. As Figure \ref{graph:qubit_connectivity_12} and \ref{graph:qubit_connectivity_40} illustrate, the vanilla QUBO model utilizing $N$ qubits can be represented as a complete graph with $N$ nodes $K_N$. There doesn't exist a natural embedding of complete graphs with number of nodes exceeding $64$ into the Chimera topology $C_{16,16,4}$. Thus, the D-wave quantum annealer can process the previously studied QUBO models which utilizes up to $64$ qubits.

Meanwhile, the proposed model significantly bounds the connectivity of qubits regardless of the number of variables used in the system of linear equations. As previously assumed, suppose the proposed model formulates a linear system in $n$ variables, with each variable represented by $m$ base-2 positive digits and $m$ base-2 negative digits. As shown in Figure \ref{graph:qubit_connectivity_12} and \ref{graph:qubit_connectivity_40}, the model can be represented as a union of $4$ complete graphs with $m$ nodes $K_m$. Experimental results suggest that the D-wave 2000Q quantum annealer succeeds in embedding $4$ copies of complete graphs with $33$ nodes $K_{33}$ into the space $C_{16,16,4}$ (Note that $K_{33}$ can be embedded into $C_{16,16,4}$ because $K_{64}$, which includes $K_{33}$, can be). This gives an explanation of why the proposed model can process $33 \times 4 = 132$ qubits instead of the $64$ qubits the previously studied QUBO models can process.
\begin{equation}
    \bigsqcup_{i=1}^4 K_{33} \hookrightarrow C_{16,16,4}
\end{equation}

Even better, the proposed model establishes a linear relation between the implementable dimension of the system of linear equations on the quantum annealer and the number of qubits on the quantum annealing device. This achievement is made possible because the connectivity of qubits is solely determined from the number of base-2 digits used to represent each variable, regardless of the number of unknown variables used in the system of linear equations. For example, given a positive number $m > 0$, suppose the quantum annealing device supports the connectivity of $2048 m^2$ qubits using the Chimera topology $C_{16m,16m,4}$. Previously researched QUBO models can process the QUBO model which utilizes $64m$ qubits \cite{BKR16, C11}. In particular, the vanilla QUBO model establishes a square-root relation between the number of qubits on the quantum annealer and the number of qubits executed by the QUBO model.
\begin{equation}
    \text{\# implementable qubits (vanilla QUBO)} \leq \sqrt{2 \times (\text{\# qubits on } C_{16m,16m,4} )}
\end{equation}
On the other hand, the new QUBO model can process $132 m^2$ qubits using the same topology. This is because the topology $C_{16m, 16m, 4}$ contains $m^2$ copies of $C_{16, 16, 4}$, each space of which can process $132$ qubits using the proposed model. Thus, we obtain:
\begin{equation} \label{equation:new_QUBO_qubits}
    \text{\# implementable qubits (proposed QUBO)} \leq \frac{33}{512} \times (\text{\# qubits on } C_{16m,16m,4} ). 
\end{equation}

Prospective quantum annealing devices are expected to support more number of qubits by enlarging the dimensions of planar grids of the Chimera topology. In other words, one shall expect linear growth in the number of variables of systems of linear equations implementable on quantum annealing devices, as shown in (\ref{equation:new_QUBO_qubits}). For example, the D-wave 7000Q quantum annealer aims to enact qubit connections on Zephyr topology $Z_m$, obtained as a quotient space of the Chimera topology $C_{2m+1,2m+1,8}$ \cite{BKR21}. We conjecture that the number of implementable unknown variables supported by the D-wave 7000Q quantum annealer will be $\frac{7}{2}$ times more than that supported by the D-wave 2000Q quantum annealer.

\subsection{Complexity}
Given the two matrices $D$ and $R$ associated to $A^T A$ from Sylvester's law of inertia, our proposed method ensures that the computational complexity of solving the QUBO model is in the order of $m n^2$. Evaluating (\ref{equation:QUBO_model_1_2}) dominates the computational burden of utilizing our proposed method. Because $b \in \mathbb{R}^n$ is a column vector of dimension $n$,
\begin{equation}
    \# \text{ calculations for computing } \sum_{i,j,k=1}^n b_k a_{k,j} r_{j,i} \left( q_{i,l}^+ - q_{i,l}^- \right) \leq 4n^2
\end{equation}
In light of the observation that the range of $l$ is between $-m$ and $m$,
\begin{equation}
    \# \text{ calculations for computing (\ref{equation:QUBO_model_1_2})} \leq 4(2m+1)n^2.
\end{equation}
Our model hence provides polynomial speedup in solving the QUBO model in comparison to previously studied quantum annealing methods, whose computational complexities are in the order of $m n^3$. \cite{BL19, J21}.

%\textcolor{red}{Additional comment.}
%약점도 같이 적어야 함.
%D and R을 임의의 matrix에 대해 구하기는 어려움.
%D and R이 이미 구해진 matrix에 대해서는 결과 계산 편함.

% 양자 QR 디컨포지션이 생각보다 느리네요 ㅠㅠ. 그래도 이거랑 연결시켜서 약점을 잘 포장하면 될거 같아요. 지금 적혀 있는것도 나쁘지는 않습니다. 아직은 조금 부족하지만 미래에 더 좋아질거다 내지는 효율적인 알고리즘이 개발되고 있다? 이런 표현은 어떨까 싶기도 하구요. 

\subsection{Limitations and relation to classical algorithms}
A central question which is left unexplored in this manuscript is determining the computational complexity of obtaining the two matrices $D$ and $R$ from Sylvester's law of inertia. Suppose that $A$ is a non-singular matrix. Then the two matrices can be obtained from the $QR$ decomposition of $A$. Indeed, we may factorize $A = QR'$ where $Q \in \mathbb{R}^{n \times n}$ is a real orthogonal matrix, and $R' \in \mathbb{R}^{n \times n}$ is an upper triangular matrix. This yields the following equation:
\begin{equation}
    (R'^{-1})^T A^T A R'^{-1} = (AR'^{-1})^T (AR'^{-1}) = Q^T Q = I.
\end{equation}
The diagonal matrix $D$ can be then chosen to be the identity matrix, and $R$ the matrix $R'^{-1}$.

One limitation of the proposed method is the potential difficulty in computing the matrices $D$ and $R$ for any given real matrix $A$. Assuming that both $D$ and $R$ are obtained, however, the QUBO model can effectively parallelize the computational process of obtaining solutions to systems of linear equations. 

Note that Sylvester's law of inertia does not necessitate the use of $QR$ factorization of $A$. One only needs to guarantee that the column vectors of the real matrix $AR \in \mathbb{R}^{n \times n}$, whose norms are the diagonal entries of $D$, are mutually orthogonal. In other words, comparative advantages that quantum algorithms possess to classical computational methods can prove to be effective in enhancing the cost efficiency of solving systems of linear equations on a quantum computer. For example, parallel computations of solutions of linear systems can be executed, which is not yet achievable on classical implementations such as QR decomposition using householder transformations. Meanwhile, QR decomposition methods can be implemented in a cost-efficient manner by utilizing intrinsic properties of quantum algorithms, including computational bases \cite{MLZ20} and synthesis of quantum circuits \cite{BBVA20}. These methods are expected to achieve polynomial speedup in computing an orthogonal basis of a vector space in comparison to previously researched classical algorithms. Thus, with prospective effective quantum implementations of Sylvester's law of inertia, we shall expect to achieve an accurate, efficient, and feasible quantum algorithm for solving systems of linear equations with large number of variables.

\section*{Declaration of Interests}
The authors have no competing interests which may have influenced the work shown in this manuscript.

\section*{Contributions}
S.W.P. conceived and designed the theoretical experiments, and the theoretical solutions. K.J. obtained the experimental results in a quantum annealer D-wave 2000. All authors wrote the paper.

%Ariv Version

%References
%\bibliography{report} % bibliography data in report.bib

\begin{thebibliography}{}
\bibitem{RI14} S.B. R{\o}nnow, T. and Isakov, S., "Evidence for quantum annealing with more than one hundred qubits," \emph{Nat. Phys} \textbf{10}, 218-224 (2014).
\bibitem{MV16} O'Malley, D. and Vesselinov, V. V., "Toq. jl: A high-level programming language for d-wave machines based on julia," \emph{2016 IEEE High Performance Extreme Computing Conference (HPEC)} 1-7. (2016)
\bibitem{BL19} Borle, A. and Lomonaco, S. J., "Analyzing the quantum annealing approach for solving linear least squares problems," \emph{International Workshop on Algorithms and Computation}, 289-301 (2019).
\bibitem{JC21} Jun, K., Conley, R., Huang, Y., Lim, H., and Yu, K., "Solving linear systems by quadratic unconstrained binary optimization on D-Wave quantum annealing device," \emph{Quantum Information Science, Sensing, and Computation XIII} \textbf{11726} 117260C (2021).
\bibitem{PLKWJ21} Park, S. W., Lee, H. J., Kim, B. C., Woo, Y., and Jun, K. "On the application of Sylvester's law of inertia to QUBO formulations for systems of linear equations," \emph{Accepted to the Proceedings of the 12th International Conference on ICT Convergence}, 1363--1367 (2021).
\bibitem{Ay62} Ayres, F., "Theory and problems of matrices," \emph{Schaum's Outline Series} 115-124 (1962).
\bibitem{S52} Sylvester, J. J., "XIX. A demonstration of the theorem that every homogeneous quadratic polynomial is reducible by real orthogonal substitutions to the form of a sum of positive and negative squares," \emph{The London, Edinburgh, and Dublin Philosophical Magazine and Journal of Science} \textbf{4} (23) 138--142 (1852).
\bibitem{J21} Jun, K. "QUBO formulations for system of linear equations" \emph{arXiv preprint arXiv:2106.10819} (2021).
\bibitem{MLZ20} Ma, G., Li, H., and Zhao, J. "Quantum QR decomposition in the computational basis," \emph{Quantum Information Processing} \textbf{19}(8), 1--16 (2020)
\bibitem{BBVA20} de Brugi{\`e}re, T. G., Baboulin, M., Valiron, B., and Allouche, C. "Quantum circuits synthesis using Householder transformations," \emph{Computer Physics Communications} \textbf{248}, 107001 (2020).
\bibitem{ZBDE20} Zbinden, S., B\''{a}rtschi, A., Djidjev, H., and Eidenbenz, S. "Embedding algorithms for quantum annealers with Chimera and Pegasus connection topologies," \emph{International Conference on High Performance Computing} 187--206 (2020)
\bibitem{BKR16} Boothby, K., King, A.D., and Roy, A. "Fast clique minor generation in Chimera qubit connectivity graphs," \emph{Quantum Information Processing} \textbf{15}, 495--508 (2016).
\bibitem{C11} Choi, V. "Minor-embedding in adiabatic quantum computation: II. Minor-universal graph design," \emph{Quantum Information Processing} \textbf{10} (3), 343--353 (2011).
\bibitem{BKR21} Boothby, K., King, A.D., and Raymond, J. "Zephyr Topology of D-Wave Quantum Processors" \emph{Technical report available at https://www.dwavesys.com/media/fawfas04/14-1056a-a\_zephyr\_topology\_of\_d-wave\_quantum\_processors.pdf} 1--18 (2021).
\end{thebibliography}
%\bibliographystyle{spiebib} % makes bibtex use spiebib.bst

\end{document}